\documentclass[twocolumn,showpacs,prd,floatfix,aps]{revtex4}
\begin{document}
\title{Instanton modifications of the bound state singularity in the Schwinger Model}
\author{Tomasz Rado\.zycki}
\affiliation{Department of Mathematical Methods in
Physics, Warsaw University, Ho\.za 74, 00-682 Warsaw,
Poland}
\email{torado@fuw.edu.pl} 

\begin{abstract}
We consider the quark-antiquark Green's function in the Schwinger Model with instanton contributions taken into account. Thanks to the fact that this function may analytically be found, we draw out singular terms, which arise due to the formation of the bound state in the theory --- the massive Schwinger boson. The principal term has a pole character. The residue in this pole contains contributions from various instanton sectors: $0,\pm 1, \pm 2$. It is shown, that the nonzero ones change the factorizability property. The formula for the residue is compared to the Bethe-Salpeter wave function found as a field amplitude. Next, it is demonstrated, that apart from polar part, there appears in the Green's function also the weak branch point singularity of the logarithmic and dilogarithmic nature. These results are not in variance with the universally adopted $S$-matrix factorization.
\end{abstract}
\maketitle

\section{Introduction}
\label{sec:intro}
The Schwinger Model (SM)~\cite{js}, describing a massless fermion field in interaction with a gauge boson in two space-time dimensions, has become a very fruitful system in field theory. Above all it possesses many of the features, one is convinced to find in the theory of strong interactions. Therefore, thanks to its relative simplicity, SM is expected to be a perfect testing layout for the various nonperturbative aspects of QCD. They involve confinement, topological sectors, instantons, condensates. Other interesting and nontrivial features one should mention, are the existence of anomaly and gauge boson nonzero mass generation without the need of introducing an auxiliary Higgs field. Naturally some of these properties are related with one another being in fact different manifestations of the same physical phenomena. 

Another aspect, that should be appreciated in SM, is the exact solvability of the equations, which allows one to perform many calculations without unpleasant approximations and additional assumptions or simplifications. One example is the quark propagator (the term `quark' means here the fundamental fermion, due to its confinement property), which has already been found by Schwinger, but also other functions stand wide open. For instance, in our previous work~\cite{trjmn}, we were able to obtain the explicit formulae for the whole set of four-point Green's functions. 

A very troublesome and long-standing question in Quantum Field Theory is the consistent description of bound states. The problem becomes particularly severe when such a state has a relativistic character and retardation effects in the interactions among constituent particles cannot be neglected. Over half a century ago, the equation for the bound state wave function --- the so called Bethe-Salpeter equation --- was formulated~\cite{bs,gml}. This equation is, however, extremely complicated. Firstly, it is a multidimensional integral equation and these are always more challenging than differential ones. Secondly, being an equation for the unknown bound state wave function, it requires, as an input, the knowledge of the nonperturbative propagators for ingredient particles and the interaction kernel between them.  Naturally no such quantities are known in realistic field theories. From the very beginning one is then doomed to extremely strong approximations. As to the calculations in field theoretical models, there are only few examples (for instance the so-called Wick-Cutkosky Model~\cite{modelbs}), where it is possible to find solutions but even in that case one is forced to simplify the equation either by neglecting relative time dependence or simplifying the interaction kernel together with one-particle propagators.

The other possibility in investigating bound states, instead of solving Bethe-Salpeter equation, is to consider four-point Green's function in the appropriate channel, and find the residue of the pole corresponding to a given bound particle~\cite{eden,mandel}. This residue is constructed just from the Bethe-Salpeter wave functions. However, knowing the full four-point function is usually as unlikely as solving exactly the Bethe-Salpeter equation itself. One should then all the more appreciate the SM, for which the {\em exact} form of the Green's function in question, was found~\cite{trjmn,lsb}. This is the main reason, why we decided to look for the bound state singularity in this model, to analyze the nature of this singularity, and to derive the formula for the Bethe-Salpeter function. The calculation performed primarily in section~\ref{sec:notop} shows the complete agreement with general considerations~\cite{eden,mandel,wc}.

A certain new aspect appears however, if one takes into account that the SM is a topologically nontrivial theory~\cite{cadam1,smil,gattr,maie,rot,gmc}. As it is well known, fermion Green's functions gain contributions from nonzero instanton sectors. This is connected with the existence in the theory of the (infinite) set of topological vacua, the superposition of which constitutes the true ground state. The expectation values for field operators in this new state, become now sums of matrix elements taken between various topological vacua. If a given operator has non-vanishing off-diagonal elements, then they have to enter into the calculation, apart from the main, diagonal contributions. Distinct topological vacua correspond to different, and topologically inequivalent, configurations of the gauge field, and the corresponding transition between them is described by the instanton field. There is a proportionality between topological index of a vacuum and the chiral charge and, therefore, off-diagonal elements may appear only for the chirality variant operators. To that category belong the products of fermion fields and, consequently, we may expect the appearance of certain new terms in the Green's functions involving quarks. The explicit calculations for the two- and four-point functions has been done in~\cite{trinst}, where these supplementary terms were explicitly given. 

Now the question arises if, and how, these contributions modify the behaviour of the two-fermion Green's function close to the bound state singularity. To this question we devote section~\ref{sec:top:subsec:1i} (one instanton contributions) and section~\ref{sec:top:subsec:2i} (two instantons contributions). Our results indicate two modifications. Firstly, we see that the residue in the polar term acquires new ingredients: the Bethe-Salpeter wave function is modified --- and this is quite obvious, since it is defined through the product of the two quark fields, which is sensitive to instanton background --- and the other term appears, which makes the residue nonfactorizable. Yet, it may still be written as a sum of several factorizable terms. Secondly, apart from the pole, we find terms, which show another type of singularity --- branch point singularity of the logarithmic and dilogarithmic nature. These contributions are small in comparison with the principal term, and in the usual bound state pole derivations~\cite{eden,mandel}, are neglected by the proof construction itself. In the SM, we happily dispose the full Green's function without being obliged to make any addition assumptions or simplifications, which usually remain beyond our control, and those terms survive.

On the other hand, the calculated instanton contributions disappear for transition amplitudes between {\em asymptotic} states and to that extent our results are in agreement with the analyticity properties of the S-matrix~\cite{jc,wzim,dio,jct,dsk,eden, iago}. For the confined particles one cannot require the cluster decomposition on the level of the Green's function. The main argument referring to the possible separation of certain subgroups of particles into the distant space-time regions, may be applied only for hadrons, but neither for individual quarks nor even $q\overline{q}$ structures, if they do not bear the complete set of quantum numbers of a meson. It is then not surprising, that the quark Green's function reveals much richer analytical structure, than the hadronic $S$-matrix does.

Finally, in section~\ref{sec:amplit} we calculate directly the Bethe-Salpeter wave function as a $q\overline{q}$ field amplitude between one-particle state of momentum $P$ and the vacuum state, and find that it corresponds to that obtained from the residue of the factorizable part of the preceding section.

\section{Basic definitions and notation}
\label{sec:def} 

The Lagrangian of the Schwinger Model, describing a fermion (quark) field $\Psi$ in interaction with a gauge boson $A^{\mu}$, has the following form:  
\begin{eqnarray} 
{\cal L}(x)&=&\overline{\Psi}(x)\left(i\gamma^{\mu}\partial_{\mu}
- g\gamma^{\mu}A_{\mu}(x)\right)\Psi (x)\nonumber\\* &-&
\frac{1}{4}F^{\mu\nu}(x)F_{\mu\nu}(x)-
\frac{\lambda}{2}\left(\partial_{\mu}A^{\mu}(x)\right)^2\; , 
\label{eq:lagr} 
\end{eqnarray} 
where $\lambda$ fixes the gauge and $g$ is the coupling constant. The space-time has one time and one space dimension. In this world, the most simply is to choose the quark field in the two-component form. In this case, the $2\times 2$ Dirac gamma matrices may have the representation given below: 
\begin{equation} 
\gamma^0=\left(\begin{array}{lr}0 & \hspace*{2ex}1 \\ 1 & 0 
\end{array}\right)\; , \;\;\;\;\; 
\gamma^1=\left(\begin{array}{lr} 0 & -1 \\ 1 & 0 
\end{array}\right)\; ,
\label{eq:gamma}
\end{equation}
$$
\gamma^5=\gamma^0\gamma^1=\left( 
\begin{array}{lr} 1 & 0 \\ 0 & -1 \end{array}\right) \; .
$$ 
For the metric tensor we choose the convention:
$$
g^{00}=-g^{11}=1\; ,
$$ 
and the antisymmetric symbol $\varepsilon^{\mu\nu}$ is defined by:
\begin{equation}
\varepsilon^{01}=-\varepsilon^{10}=1\; ,\;\;\;\;\; 
\varepsilon^{00}=\varepsilon^{11}=0\; .
\label{eq:epsil} 
\end{equation}
All the Green's functions may be obtained by differentiating over external currents of the 
generating functional, which is, as usually, defined by the Feynman path integral:
\begin{equation}
Z[\eta,\overline{\eta},J]=\int{\cal D}\Psi{\cal D}\overline{\Psi}{\cal 
D}Ae^{i \int d^2x \left[{\cal 
L}+\overline{\eta}\Psi+\overline{\Psi}\eta+J^{\mu}A_{\mu}\right]}\; ,
\label{eq:genfunc}
\end{equation}
As we have already written in the Introduction, this theory is topologically nontrivial. The true vacuum, which is usually called the $\theta$-vacuum, becomes a superposition of the infinite set of degenerate vacua, bearing different topological indices $n$~\cite{rajar,huang}:
\begin{equation}
|\theta>=\sum_{n=-\infty}^{\infty}e^{in\theta}|n>\; .
\label{eq:vacuum}
\end{equation}
In this new vacuum, the generating functional~(\ref{eq:genfunc}), which constitutes the Feynman form of $<\theta |\theta>_{\eta ,\overline{\eta}, J}$, breaks into pieces:
\begin{equation}
Z[\eta ,\overline{\eta}, J]=\sum_{k=-
\infty}^{\infty}e^{ik\theta}Z^{(k)}[\eta ,\overline{\eta}, J]\; .
\label{eq:gener}
\end{equation}
In this formula --- contrary to~(\ref{eq:vacuum}) --- the summation runs over instanton number, i.e. $k$ corresponds to the change of the topological indices in the two vacua in the expression: $<\theta |\theta>_{\eta ,\overline{\eta}, J}$. Henceforth we will use the common symbol $|0>$, for denoting the true vacuum. In each of $Z^{(k)}$'s the functional integration with respect to the gauge field $A^{\mu}$, is restricted to a specific class of configurations bearing the Pontryagin index $k$.

Now the $q\overline{q}$ Green's function may be represented as the appropriate functional derivative of $Z$:
\begin{eqnarray}
&&G_{ab,cd}(x_1,x_2;x_3,x_4)=\label{eq:Greendef}\\*
&&\;\; =<0|T(\Psi_a(x_1)\Psi_b(x_2)\overline{\Psi}_c(x_3)\overline{\Psi}_d(x_4))|0>=\nonumber\\*
&&\;\; =N\left.\frac{\delta^4Z}{\delta\overline{\eta}_a(x_1)\delta 
\overline{\eta}_b(x_2)\delta\eta_c(x_3)\delta\eta_d(x_4)}\right|_{\eta ,\overline{\eta}, J=0}\; ,\nonumber
\end{eqnarray}
where $N$ is a constant ensuring that $<0 |0>_{\eta ,\overline{\eta}, J=0}=1$. In the following, we are only interested in singularities that may be associated with bound states, and hence, throughout this work, by $G$ we understand only the connected part of this function. 

Due to the fact, that $Z$ expands in the sum of $Z^{(k)}$'s, the same refers also to the Green's function. However, this sum is now limited only to the low values of $k$. Contrary to the expression $\exp\left\{i \int d^2x \left[\overline{\eta}\Psi+\overline{\Psi}\eta\right]\right\}$, which contains arbitrary high powers of $\Psi$ and $\overline{\Psi}$ and, therefore, contributes to all sectors in~(\ref{eq:gener}), the four-fermion function gets terms with at most $k=\pm 2$. This is due to the fact, that four fermion fields can change the chirality of the state by (at most) 4, and the topological vacuum index is a half of the chiral charge~\cite{rajar}. In the following sections we investigate in detail all possible terms.

\section{Quark-antiquark bound state without topological contributions}
\label{sec:notop}
In this section we will concentrate on the 0-instanton term of the function $G_{ab,cd}(x_1,x_2;x_3,x_4)$. It is a great merit of the Schwinger Model, that this function may be found exactly, in an explicit and analytic way. The details of our calculation have already been given in~\cite{trjmn} and we will not elaborate on them here, but instead recall only the final result: 
\begin{widetext}
\begin{eqnarray}
G^{(0)}_{ab;cd}&&\!\!\!\!\!(x_1,x_2;x_3,x_4)=\nonumber\\*
&&=\frac{1}{2}\{S_{ac}(x_1-x_3)S_{bd}(x_2-x_4)
+[S(x_1-x_3)\gamma^5]_{ac}[S(x_2-x_4)\gamma^5]_{bd}\}\exp\{ig^2[\beta(x_1-x_2)-\beta(x_1-x_4)\nonumber\\*
&&-\beta(x_2-x_3)+\beta(x_3-x_4)]\}+\frac{1}{2}\{S_{ac}(x_1-x_3)S_{bd}(x_2-x_4)- [S(x_1-x_3)\gamma^5
]_{ac}[S(x_2-x_4)\gamma^5]_{bd}\}\nonumber\\*
&&\times\exp\{-ig^2[\beta(x_1-x_2)-\beta(x_1-x_4)-\beta(x_2-x_3)
+\beta(x_3-x_4))]\}-\left\{\begin{array}{ccc} x_3 & \leftrightarrow & x_4\\ c &
\leftrightarrow & d 
\end{array}\right\}\; ,
\label{eq:Green's0}
\end{eqnarray}
\end{widetext}
where the index $(0)$ refers to the instanton number. The function $\beta(z)$ is defined through the two-dimensional momentum integral:
\begin{equation}
\beta(z)=\int\frac{d^2p}{(2\pi)^2}\frac{\left(1-e^{ipz}\right)}{(p^2-g^2/\pi
+i\epsilon)(p^2+i\epsilon)}\; ,
\label{eq:beta}
\end{equation}
and may be expressed in terms of MacDonald or Hankel function, depending on whether $z$ is space- or time-like, but the above representation is more suitable for us. 
We recall that in 2D the coupling constant $g$ is dimensionfull and the quantity $\mu^2=g^2/\pi$ plays a role of the mass squared of the so called Schwinger boson -- a $q\overline{q}$ bound state we are interested in. 

The object $S(x)$ in~(\ref{eq:Green's0}) is the full quark propagator, which has the known form:
\begin{equation}  
S(x)={\cal S}_0(x)\exp\left[-ig^2\beta(x)\right]\; ,  
\label{props} 
\end{equation}
with ${\cal S}_0(x)=-\frac{1}{2\pi}\frac{\not x}{x^2-i\varepsilon}$ being the free propagator. The term `full' used above, means here `full in the 0-instanton sector', i.e. it is given in the form of the original Schwinger works~\cite{js}. However $S$ -- bilinear in $\Psi$'s --- must later get contributions from the nontrivial topological configurations~\cite{trinst}. 

Now we will concentrate on the structure of the $q\overline{q}$ function~(\ref{eq:Green's0}) and try to reveal, whether it has a pole in the $t$-channel. To this goal we introduce new space-time variables: $X$, $Y$, $x$ and $y$, defined in the following way:
\begin{eqnarray}
X=\frac{1}{2}(x_1+x_3)\; ,&&\;\;\;\; x=x_1-x_3\; ,\nonumber\\*
Y=\frac{1}{2}(x_2+x_4)\; ,&&\;\;\;\; y=x_2-x_4\; .\label{eq:coord}
\end{eqnarray}
Thanks to the translational invariance, which manifests itself through the dependence of $G$ on $x_i-x_j$'s only, we have three independent variables: $Z=Y-X$, $x$ and $y$. From the four terms present in~(\ref{eq:Green's0}), the two explicitly written are of interest for us, since the two other, obtained by the antisymmetrization, contribute to the singularities not in $t$, but eventually in the $u$-channel. First of the two interesting terms, denoted by $I$, takes the form:
\begin{eqnarray}
&&G^{(0)I}(Z;x,y)\!\!=\!\!\frac{1}{2}\{S(x)\otimes S(y)
+[S(x)\gamma^5]\otimes [S(y)\gamma^5]\}\nonumber\\*
&&\times\exp\{ig^2[\beta(-Z+(x-y)/2)-\beta(-Z+(x+y)/2))\nonumber\\*
&&-\beta(-Z-(x+y)/2))+\beta(-Z-(x-y)/2))]\}\; ,\label{eq:g01}
\end{eqnarray}
where we made use of the evenness of $\beta$: $\beta(x)=\beta(-x)$. For brevity we omitted  the spinor indices, and used the tensor product notation, which should not cause any confusion. The pole corresponding to the bound state should be found in the complex plane of $P^2$, where $P$ is the two-momentum canonically conjugated to $Z$. The matrical part depends only on the relative coordinates $x$ and $y$ and enters into the residue of this pole, but does not influence its position. Let us then perform the Fourier transform over $Z$ of the exponent containing $\beta$ functions. Making use of~(\ref{eq:beta}) we my write the quantity in question as:
\begin{equation}
F^I(P,x,y)=\int d^2Z e^{iPZ}e^{-4ig^2\int\frac{d^2p}{(2\pi )^2}e^{-ipZ}I_1(p,x,y)}\; ,
\label{eq:FI}
\end{equation}
where $I_1(p,x,y)=\frac{\sin(px/2)\sin(py/2)}{(p^2-\mu^2+i\epsilon)(p^2+i\epsilon)}$. Expanding the exponent in series, we can execute the $Z$ integration obtaining:
\begin{equation}
F^I(P,x,y)=\sum_{n=1}^{\infty}\frac{(-4ig^2)^n}{n!}I_n(P,x,y)+(2\pi)^2\delta^{(2)}(P)\; ,
\label{eq:FIexp}
\end{equation}
The $I_n$'s satisfy the following recurrent relation:
\begin{eqnarray}
I_n&&\!\!\!\!\!(q,x,y)=\label{eq:rec}\\*
&&=\int\frac{d^2p}{(2\pi )^2} \frac{\sin(px/2)\sin(py/2)}{(p^2-\mu^2+i\epsilon)(p^2+i\epsilon)}I_{n-1}(q-p,x,y)\nonumber\; .
\end{eqnarray}
It is obvious that $I_1(p,x,y)$ has a pole at $p^2=\mu^2$. The other singularity is at $p^2=0$. But what about terms of higher $n$? Do they alter the behaviour at $p^2=\mu^2$? Let us first construct $I_2(p,x,y)$. According to~(\ref{eq:rec}) it has the form:
\begin{eqnarray}
I_2&&\!\!\!\!\!(q,x,y)=\label{eq:I2}\\*
&&=\int\frac{d^2p}{(2\pi )^2} \frac{\sin(px/2)\sin(py/2)}{(p^2-\mu^2+i\epsilon)(p^2+i\epsilon)}I_1(q-p,x,y)\nonumber\; .
\end{eqnarray}
The possible singularities in $q^2$, exhibited in $I_2$, can come from the singularities of the integrand functions. There are four of them: obliviously $p^2=0$, $p^2=\mu^2$, and the other two originating from $I_1$. These are: $(p-q)^2=0$ and $(p-q)^2=\mu^2$. It is not possible to satisfy all these four or even three conditions simultaneously, so the singular points may appear only as a result of the merging of two singularities. In accordance with Landau procedure~\cite{landau,eden,elop}, we introduce two parameters ($\alpha$ and $\beta$) satisfying $\alpha,\beta \geq 0$ and $\alpha^2+\beta^2\neq 0$. With the assumption $\alpha p+\beta(p-q)=0$, the following possibilities occur:
\begin{enumerate}
\item $\alpha p^2=0$ and $\beta (q-p)^2=0$. If one of the parameters equals zero, then we get $p=0$ or $p-q=0$ and the eventual singularity is cancelled by the sine functions present in the numerator in~(\ref{eq:I2}), either by that explicitly written or by the one hidden in the formula for $I_1$. If both parameters are nonzero, on can easily obtain $q^2=\left(\frac{\alpha+\beta}{\beta}\right)^2p^2=0$. Thus, there may only arise a contribution to the singularity at $q^2=0$. It cannot, however, be worse than that in $I_1$.
\item $\alpha (p^2-\mu^2)=0$ and $\beta(p-q)^2=0$. If $\alpha=0$, we have $q=p$ and, in general, one might get a contribution to the pole at $q^2=\mu^2$ (actually it would become a branch point), but it does not happen because of the sine functions becoming zero for $q-p=0$. If $\beta=0$, we obtain $p=0$, which stays in contradiction with the assumption $p^2=\mu^2$. For $\alpha,\beta\neq 0$, we find $\alpha^2p^2=\beta^2(p-q)^2$, which is incompatible with $p^2=\mu^2$ and $(p-q)^2=0$.
\item $\alpha p^2=0$ and $\beta((p-q)^2-\mu^2)=0$. This case is almost identical with the previous one and need not be separately considered.
\item $\alpha (p^2-\mu^2)=0$ and $\beta[(p-q)^2-\mu^2]=0$. If one of the parameters is zero, then we find a contradiction of $\alpha p+\beta(p-q)=0$, with either $p^2=\mu^2$ or $(p-q)^2=\mu^2$. What remains is to assume $\alpha,\beta\neq 0$. If so, we have $\alpha^2p^2=\beta^2(p-q)^2$ and after cancelling the arising factors $\mu^2$ on both sides, we get $\alpha=\beta$. This leads to $q=2p$ and a singularity emerges now at $q^2=4\mu^2$. This is the two-particle singularity.
\end{enumerate}
Summarizing, we observe that in $I_2$ the pole at $\mu^2$ disappears and a new singularity arises at $(2\mu)^2$. Now we are in a position to carry out a proof by induction: the assumed singularities under the integral in~(\ref{eq:rec}) are: $p^2=0$, $p^2=\mu^2$, $(p-q)^2=0$ and $(p-q)^2=((n-1)\mu)^2$. We have again four possibilities, the analysis of which is identical as above, except the last point, where we now get the equation $\alpha^2=(n-1)^2\beta^2$. This leads to $q=np$ and $q^2=n^2\mu^2$. Hence, the conclusion is, that the singularity corresponding to $n$ intermediate particles ($q^2=n^2\mu^2$) is present only in $I_n$. This means, that the pole at $\mu^2$ may be simply picked up from the first term of the expansion~(\ref{eq:FIexp}).

Now we need to apply the same to the second expression in the Green's function~(\ref{eq:Green's0}). Again we take out the exponent and perform the Fourier transform, the resulting expression being called $F^{II}(P,x,y)$. Happily both exponents differ only by the sign, so we are able to immediately write down the final result. The total contribution (i.e. $I+II$) to the bound state pole in $G^{(0)}$ is:
\begin{eqnarray}
G_{b.s.}^{(0)I}(P;x,y)=&&-4i\pi\left[S(x)\gamma^5\right]\otimes \left[S(y)\gamma^5\right]\nonumber\\*
&&\times\frac{\sin (Px/2) \sin (Py/2)}{P^2-\mu^2+i\epsilon}\; ,
\label{eq:pol0}
\end{eqnarray}
We see the perfect factorization of the residue, and the quantity: \begin{equation}
\Phi_P(x)=-2\sqrt{\pi}S(x)\gamma^5\sin (Px/2)\; ,
\label{eq:phi0}
\end{equation}
plays a role of the Bethe-Salpeter amplitude~\cite{bs,gml,mandel} or the wave function. The minus sign is chosen for later convenience. Note, that this quantity is found explicitly and exactly, without any simplification referring neither to the relative time $x^0$ nor to the structure on the interaction kernel. It is an essential point, since even in the simplified model theories, one usually is limited to the case of the so called `equal time wave function' (i.e. defined only on the hypersurface $x^0=0$ in the center of mass frame), which neglects retardation effects and consequently leads to the nonrelativistic bound state, or ladder approximation for the interaction kernel. On the other hand, the function~(\ref{eq:phi0}) is a fully relativistic object.

\section{Topological contributions to the quark-antiquark bound state}
\label{sec:top}
As it has been mentioned, the vacuum in the Schwinger Model is nontrivial: it is a superposition~(\ref{eq:vacuum}) of vacua corresponding to various topological numbers. As a consequence, all chirality changing operators --- and to that class the product of fermion fields belongs --- which have nonvanishing matrix elements between different topological vacua, acquire contributions from nonzero instanton sectors. We recall here, that if the theory contains massless fermions (at least one) --- and this is just the case of the Schwinger Model --- the amplitude of the spontaneous tunnelling between 
different topological vacua vanishes~\cite{thoo,rajar}. Mathematically it is 
expressed through zero value of the euclidean Dirac operator determinant: 
$\det [i\not\!\! \partial - e\not\!\!\! A]$, because of the appearance of
zero eigenvalues when $A^\mu$ bears the instanton 
index~\cite{thoo,as,kis,ans,huang}. The functional integral over 
fermionic degrees of freedom, corresponding to the transition amplitude in 
question, is then zero and the tunnelling phenomenon disappears. This does not mean, however, that the notion of $\theta$-vacuum is useless, since topological vacua (contrary to the $\theta$-vacuum) violate cluster property~\cite{cdg}. We are the fated to use~(\ref{eq:vacuum}) and must put up with modification in the fermion Green's functions

The contributions to the four-point function were already found in our previous work~\cite{trinst}. Below we will concentrate on the eventual bound state pole in the formulae.

\subsection{1-instanton sector}
\label{sec:top:subsec:1i}
By virtue on the Atiyah-Singer index theorem~\cite{as,thoo,kis,ans,huang}, the massless Dirac operator in the external field bearing nonzero instanton number $k$ has zero modes. 
In the present section we consider the case of $k=\pm 1$, for which there is only one such mode. Its detailed form may be found in the literature~\cite{cadam1,maie,jaye}) and we will not recall it here. This mode gives now its contribution while calculating the Feynman path integral~(\ref{eq:genfunc}). Since it corresponds to the zero eigenvalue, it disappears from the quadratic part of the Lagrangian and remains only in source terms. Due to the Grassman character of the quark fields, of the four differentiations in~(\ref{eq:Greendef}), two must be saturated by sources multiplying zero modes. The other two are performed in a common way. The result for the four-point function, with both $k=+1$ and $k=-1$ instanton numbers taken into account, obtained in~\cite{trinst}, is:
\begin{widetext}
\begin{eqnarray}
G^{(1)}(x_1,x_2;x_3,x_4) =&&-\frac{ig}{8\pi^{3/2}}e^{\gamma_E}(\cos\theta-
i\gamma^5\sin\theta)\otimes{\cal S}_0(x_2-x_4)\big[(\openone\otimes\openone - 
\gamma^5\otimes\gamma^5)\nonumber\\*
&&\times e^{ig^2[\beta(x_1-x_3) - 
\beta(x_2-x_4) - \beta(x_1-x_4) + \beta(x_1-x_2) + \beta(x_2-x_3) - 
\beta(x_3-x_4)]} +(\openone\otimes\openone + 
\gamma^5\otimes\gamma^5) \label{eq:Green's1}\\*
&&\times e^{ig^2[\beta(x_1-x_3) - 
\beta(x_2-x_4) + \beta(x_1-x_4) - \beta(x_1-x_2) - \beta(x_2-x_3) + 
\beta(x_3-x_4)]}\big]+\; {\rm antisymmetrization}\; . \nonumber
\end{eqnarray}
\end{widetext}
The antisymmetrization runs here both over `initial', i.e. $x_1$ and $x_2$ and `final' i.e. $x_3$ and $x_4$ (accompanied by the appropriate spinor indices change) variables in $G(x_1,x_2;x_3,x_4)$. Thus~(\ref{eq:Green's1}) gets three more terms, which are not explicitly written here for brevity. One of them (with the simultaneous substitution $1\leftrightarrow 2$ and $3\leftrightarrow 4$) again contributes to the pole in the $t$-channel, the other two only in the $u$-channel and will be omitted. Introducing center of mass and relative variables~(\ref{eq:coord}), similarly as it was done in section~\ref{sec:notop}, we obtain for the first term in~(\ref{eq:Green's1}):
\begin{widetext}
\begin{eqnarray}
&&G^{(1)Ia}(Z;x,y)=-\frac{ig}{8\pi^{3/2}}e^{\gamma_E}\big[(\cos\theta-
i\gamma^5\sin\theta)e^{ig^2\beta(x)}\otimes{\cal S}(y)\big](\openone\otimes\openone - 
\gamma^5\otimes\gamma^5)\label{eq:g11}\\*
&&\;\;\;\;\;\;\times\exp\{ig^2[-\beta(-Z+(x+y)/2)+\beta(-Z+(x-y)/2))+\beta(-Z-(x+y)/2))-\beta(-Z-(x-y)/2))]\}\; ,\nonumber
\end{eqnarray}
\end{widetext}
The last exponent in the above, {\em per analogiam} to~~(\ref{eq:FI}), may now be written in the form:
\begin{equation}
e^{4g^2\int\frac{d^2p}{(2\pi )^2}e^{-ipZ}J_1(p,x,y)}\; ,
\label{eq:ex1}
\end{equation}
where $J_1(p,x,y)=\frac{\cos(px/2)\sin(py/2)}{(p^2-\mu^2+i\epsilon)(p^2+i\epsilon)}$. Now we might introduce $J_n$'s and consider the recurrent relation similar to~(\ref{eq:rec}), but it is unnecessary.
Due to the presence of the sine function in the numerator, we can directly apply the analysis of section~\ref{sec:notop} and come to the result that the contribution to the bound state comes only from the first term of the series for the exponent~(\ref{eq:ex1}). As to the second term (in our nomenclature it would be called $Ib$) in~(\ref{eq:Green's1}) the exponent has only the sign inverted and need not be elaborated on. Their joint effect is then simply to cancel $\openone\otimes\openone$ in~(\ref{eq:Green's1}) and to double $\gamma^5\otimes\gamma^5$. Their contribution to the bound state is therefore:
\begin{eqnarray}
G_{b.s.}^{(1)I}(P;x,y)=&&i\mu e^{\gamma_E}\big[e^{ig^2\beta(x)}e^{-i\gamma^5\theta}\gamma^5\big]\otimes \big[S(y)\gamma^5\big]\nonumber\\*
&&\times\frac{\cos (Px/2) \sin (Py/2)}{P^2-\mu^2+i\epsilon}\; ,
\label{eq:pol1}
\end{eqnarray}
where, instead of $\cos\theta-
i\gamma^5\sin\theta$, we have written $e^{-i\gamma^5\theta}$. The antisymmetrized term, which is left so far, has an analogous form, but with $x$ and $y$ swopped, and tensor products inverted. Our final result in this sector is:
\begin{widetext}
\begin{eqnarray}
G_{b.s.}^{(1)}(P;x,y)=&&\frac{i\mu e^{\gamma_E}}{P^2-\mu^2+i\epsilon}\big\{\big[e^{ig^2\beta(x)}e^{-i\gamma^5\theta}\gamma^5\big]\otimes \big[S(y)\gamma^5\big]\nonumber\\*
&&\times\cos (Px/2) \sin (Py/2)
+\big[S(x)\gamma^5\big]\otimes \big[e^{ig^2\beta(y)}e^{-i\gamma^5\theta}\gamma^5\big]\sin (Px/2) \cos (Py/2)\big\}\; .
\label{eq:pol2}
\end{eqnarray}
\end{widetext}
Thanks to the quark field being massless, which results in the chiral invariance of the initial Lagrangian, the dependence on the parameter $\theta$ may be gauged away, so one might simply put $\theta=0$. Comparing the two obtained equations~(\ref{eq:pol0}) and~(\ref{eq:pol2}) we see that introducing a new Bethe-Salpeter amplitude in the form:
\begin{eqnarray}
\Phi_P(x)=&&-2\sqrt{\pi}S(x)\gamma^5\sin (Px/2)\label{eq:phi1}\\*
&&+\frac{\mu}{2\sqrt{\pi}}e^{\gamma_E}e^{ig^2\beta(x)}e^{-i\gamma^5\theta}\gamma^5\cos(Px/2)\; ,\nonumber
\end{eqnarray}
the residue factorization in the pole will be restored i.e. one will have:
\begin{equation}
G_{b.s.}(P;x,y)\sim \frac{\Phi_P(x)\otimes\Phi_P(y)}{P^2-\mu^2+i\epsilon}
\label{eq:factor}
\end{equation}
if the 2-instanton sector contributes to the bound state a term (with the appropriate constant coefficient): 
\begin{eqnarray}
[e^{ig^2\beta(x)}e^{-i\gamma^5\theta}\gamma^5]&\otimes &[e^{ig^2\beta(y)}e^{-i\gamma^5\theta}\gamma^5] \nonumber\\*
&&\!\!\!\!\!\!\!\!\!\!\!\!\!\!\!\!\!\!\!\!\!\!\!\!\!\!\!\!\!\!\!\!\!\!\!\times\cos (Px/2) \cos (Py/2)\; .\nonumber
\end{eqnarray}
As we shall see in the following section, this is actually the case, but it contributes also other terms that modify the factorization and even change the character of the singularity at $P^2=\mu^2$.

\subsection{2-instanton sector}
\label{sec:top:subsec:2i}
In the 2-instanton sector there are two zero modes of the the Dirac operator~\cite{thoo,as,cadam1,maie,jaye}. Now all of the differentiations in~(\ref{eq:Greendef}) are saturated with sources standing with zero modes.
The 2-instanton contribution to the Green's function, which was evaluated in~\cite{trinst}, has the form:
\begin{widetext}
\begin{eqnarray}
G^{(2)}(x_1,x_2;x_3,x_4) =&& \!\!\!\!\!\! -
\frac{g^4}{256\pi^4}e^{4\gamma_E}\big[e^{2i\theta}\left(\openone - 
\gamma^5\right)\otimes\left(\openone - \gamma^5\right)(x_2^0+x_2^1) 
\label{Green's2}\\*
&&\!\!\!\!\!\! \times(-x_4^0+x_4^1) + e^{-2i\theta}\left(\openone + 
\gamma^5\right)\otimes\left(\openone + \gamma^5\right)(-x_2^0+x_2^1) 
(x_4^0+x_4^1)\big]\exp\big\{ig^2\big[\beta(x_1-x_4) \nonumber\\* 
&&\!\!\!\!\!\! +\beta(x_2-x_3) +\beta(x_1-x_2) + 
\beta(x_3-x_4) + \beta(x_1-x_3) + \beta(x_2-x_4)\big]\big\} + \; {\rm 
antisymmetrization}\; . \nonumber
\end{eqnarray}

Using again the variables $Z$, $x$ and $y$, performing the required antisymmetrization, and replacing $\beta$ functions with integral formula~(\ref{eq:beta}), we obtain, with the additional use of the identity $e^{-i\theta\gamma^5}(\openone\pm \gamma^5)= e^{\mp i\theta}(\openone\pm \gamma^5)$:

\begin{eqnarray}
G^{(2)}(Z;x,y)=&&-\frac{\mu^4}{64\pi^2}e^{4\gamma_E}\big[(e^{ig^2\beta(x)}e^{-i\gamma^5\theta})\otimes (e^{ig^2\beta(y)}e^{-i\gamma^5\theta})\big]
\times\big[(Z^2-\frac{1}{4}(x-y)^2)\gamma^5\otimes \gamma^5 \nonumber\\*
&&+\frac{1}{2}(Z^2-\frac{1}{4}(x-y)^2)\big((\openone-\gamma^5)\otimes (\openone+\gamma^5)+(\openone+\gamma^5)\otimes (\openone-\gamma^5)\big)
\label{eq:gr2}\\*
&&-\frac{1}{2}\varepsilon^{\alpha\beta}Z_\alpha(x-y)_\beta(\openone\otimes\gamma^5+\gamma^5\otimes\openone)\big]\,
\exp\big[4\pi i \mu^2 \int\frac{d^2p}{(2\pi)^2}\frac{1-e^{-ipZ}\cos(px/2)\cos(py/2)}{(p^2+i\epsilon)(p^2-\mu^2+i\epsilon)}\big]\; ,\nonumber
\end{eqnarray}
\end{widetext}
with $\varepsilon^{\alpha\beta}$ being the antisymmetric symbol defined in~(\ref{eq:epsil}). Now we will show that the first term (i.e. $Z^2$ times $\gamma^5\otimes\gamma^5$) gives just the contribution needed to restore the factorization~(\ref{eq:factor}). 

One could suspect that there might be some cancellations, among various expressions in the formula~(\ref{eq:gr2}). Below we shall argue that the cancellation that would lead to complete disappearance of any term, is not possible due to their independent matrical structure. Only cancellations among a specific matrix elements --- but not matrices as a whole --- are admissible. Consider namely the expression:
\begin{eqnarray}
a\gamma^5\otimes\gamma^5 &+&b[(1-\gamma^5)\otimes (1+\gamma^5)+(1+\gamma^5)\otimes (1-\gamma^5)]\nonumber\\*
&+& c(\openone\otimes\gamma^5+\gamma^5\otimes\openone)
\label{eq:matind}
\end{eqnarray}
and multiply it firstly by $(\openone+\gamma^5)\otimes (\openone+\gamma^5)$ and secondly by $(\openone-\gamma^5)\otimes (\openone-\gamma^5)$. We obtain $(a+2c)(\openone+\gamma^5)\otimes (\openone+\gamma^5)+b\cdot 0$ and $(a-2c)(\openone-\gamma^5)\otimes (\openone-\gamma^5)+b\cdot 0$ respectively. On the other hand if we take the double trace (i.e over first matrix and over second matrix in the tensor product indices), we get $a\cdot 0+8\cdot b +0\cdot c$. This shows that the three tensors listed in~(\ref{eq:matind}) are really independent.

Let us observe now, that the quantity in the last exponent of~(\ref{eq:gr2}), after rearranging the terms under the integral (the integrand function will be called below $M$) and partially executing it, may be rewritten in the following way:
\begin{eqnarray}
&&4\pi i \mu^2 \int\frac{d^2p}{(2\pi)^2}M(Z,p,x,y)
=-2\gamma_E -\ln(-\mu^2Z^2/4)\nonumber\\*
&&\;\;\;\;\; -4\pi i\int\frac{d^2p}{(2\pi)^2}e^{-ipZ}\frac{1}{p^2-\mu^2+i\epsilon}\label{eq:exa}\\*
&&\;\;\;\;\; +4\pi i\mu^2\int\frac{d^2p}{(2\pi)^2}e^{-ipZ}\frac{1-\cos(px/2)\cos(py/2)}{(p^2+i\epsilon)(p^2-\mu^2+i\epsilon)}\; ,\nonumber
\end{eqnarray}
If we expand, as before, the exponent of the last two terms and take the Fourier transform over $Z$, it becomes obvious that both of them contribute to the pole at $P^2=\mu^2$. The further analysis is again identical to that of Sec.~\ref{sec:notop}, except for the difference, that now there is additional cancellation between the integrals in~(\ref{eq:exa}). Finally all the factors combine together in~(\ref{eq:gr2}) and we obtain the expected term:
\begin{eqnarray}
&&\frac{-i\mu^2}{4\pi}e^{2\gamma_E}(e^{ig^2\beta(x)}e^{-i\theta\gamma^5}\gamma^5)\otimes (e^{ig^2\beta(y)}e^{-i\theta\gamma^5}\gamma^5)\nonumber\\*
&&\;\;\;\;\;\;\;\;\;\;\;\;\;\;\;\;\;\;\times\frac{\cos(Px/2)\cos(Py/2)}{P^2-\mu^2+i\epsilon}\; .\label{eq:pol3}
\end{eqnarray}
This is exactly what we needed for~(\ref{eq:factor}). This is not, however, the whole story. The similar contribution comes from the second matrical term in~(\ref{eq:gr2}), which also contains $Z^2$. Since we have already saturated~(\ref{eq:factor}) this supplementary part violates the factorization. It has the form:
\begin{eqnarray}
&&\frac{-i\mu^2}{8\pi}e^{2\gamma_E}(e^{ig^2\beta(x)}e^{-i\theta\gamma^5}\otimes e^{ig^2\beta(y)}e^{-i\theta\gamma^5})\nonumber\\*
&&\;\;\;\;\;\;\;\;\;\;\;\;\times[(\openone-\gamma^5)\otimes (\openone+\gamma^5)+(\openone+\gamma^5)\otimes (\openone-\gamma^5)]\nonumber\\*
&&\;\;\;\;\;\;\;\;\;\;\;\;\times\frac{\cos(Px/2)\cos(Py/2)}{P^2-\mu^2+i\epsilon}\; .\label{eq:nonf}
\end{eqnarray}
Due to the above mentioned lack of cancellations among independent matrix structures, this term must survive. Moreover it cannot cancel with the $(x-y)^2$ term with the same matrical coefficient since, contrary to~(\ref{eq:nonf}), the latter disappears for $x=y$. Besides, it does not contribute to the bound state pole at all. This can be easily argued as follows. After rearrangements identical to~(\ref{eq:exa}) one gets the additional $1/Z^2$ coming from the logarithm in the exponent. This can be neutralized, if we act with the two-dimensional d'Alambert operator in variable $P$. When this is done, we obtain a first order pole at $P^2=\mu^2$. Now the question is: given a function $f$ satisfying: $\partial^2_Pf(P)\sim\frac{1}{P^2-\mu^2}$. Does $f$ also display a pole? The answer is negative. It may easily be found that $f(P)\sim {\rm Li}_2(P^2/\mu^2)$ and dilogarithm has a limit $\pi^2/6$ when its argument goes to unity. Therefore it may be neglected in comparison with the polar term. One should emphasize, however, that although dilogarithmic function has a well defined limit, it is not an analytic function in this point. It is a branch point and the function has a cut on the half line $[\mu^2,\infty[$. The presence of $\cos(Px/2)\cos(Py/2)$ does not change the conclusions, since cosine is an analytic function, and may only make things better. On the other hand for $x=y=0$ cosines can be replaced with unities and a branch point arises. Such terms are usually lost in the common approach, where one fishes out only the polar contribution of the bound state.

For the term proportional to $\varepsilon^{\alpha\beta}Z_\alpha(x-y)_\beta$, similar argumentation is applicable, with a difference that we now have to do with logarithmic function, which again may be neglected if compared to the polar term. Hence, as we see, due to the topological background, not only nonfactorization appears, but the character of the singularity in $P^2=\mu^2$ is slightly changed. 

These results do not, however, contradict the $S$-matrix pole factorization~\cite{jc,wzim,dio,jct,dsk}. The $S$-matrix describes the transition amplitudes among physical asymptotic states. In the Schwinger Model these are the Fock states of Schwinger bosons. Consequently we are limited only to the current-current Green's functions and this require putting $x,y\rightarrow 0$, and taking the trace over spinor indices with $\gamma^\mu\otimes \gamma^\nu$ matrices. Since the trace of an odd number of $\gamma$'s is zero, it is easy to observe that the only contribution to the pole comes now from~(\ref{eq:pol0}) and factorization reappears~\cite{cadam1}. This is what should be expected since the vector current does not couple to instantons.

Summarizing, that part of the full $q\overline{q}$ Green's function, in which one is limited to the polar terms, has the form:
\begin{eqnarray}
&&iG_{b.s.}(P;x,y)=\frac{1}{P^2-\mu^2+i\epsilon}[\Phi_P(x)\otimes\Phi_P(y)\label{eq:poltot}\\*
&&\;\;\;\;\;\;\;+\Psi^{(-)}_P(x)\otimes\Psi^{(+)}_P(y)+\Psi^{(+)}_P(x)\otimes\Psi^{(-)}_P(y)]\; ,\nonumber
\end{eqnarray}
with 
\begin{equation}
\Psi^{(\pm)}_P(x)=\frac{\mu}{2\sqrt{2\pi}}e^{\gamma_E}e^{ig^2\beta(x)}e^{-i\theta\gamma^5}(\openone\pm\gamma^5)\; ,
\label{eq:ppm}
\end{equation}
and $\Phi_P(x)$ given by~(\ref{eq:phi1}).

\section{Wave function as a field amplitude}
\label{sec:amplit}
The question arises, how the obtained results harmonize with the Bethe-Salpeter wave function defined as an amplitude of fields. Fortunately in the Schwinger Model we have all tools, which allow us to find it explicitly even in the nontrivial topological sectors. Let us then consider the amplitude in question: 
\begin{equation}
\chi_P(x_1,x_2)=<0|T(\Psi(x_1)\overline{\Psi}(x_2)|P>\; .
\label{eq:bsampl}
\end{equation} 
The calculation of this quantity turns out to be very simple, if one already knows the expression for the full quark propagator in the instanton background~\cite{trinst}:
\begin{equation}
S_{\rm full}(x)=S(x)+\frac{ie}{4\pi^{3/2}}\left(\cos\theta-i\gamma^5\sin\theta\right)e^{\gamma_E 
+ie^2\beta(x)}\; .
\label{eq:sfull}
\end{equation}
The state $|P>$ is the one Schwinger boson state of two-momentum $P$, and~(\ref{eq:bsampl}) is expressible, thanks to the formulae of the LSZ formalism, through the vertex function.
It is known that, thanks to the two gauge symmetries ($U(1)$ and $U_A(1)$), the vertex function may in turn entirely be constructed from the quark propagator. This is a miracle of the Schwinger Model. One then gets:
\begin{eqnarray}
\chi_P(x_1,x_2)=&&\!\! i\sqrt{\pi}\big[e^{-iPx_1}\gamma^5 S(x_1-x_2)\nonumber\\*
&&\!\! +e^{-iPx_2}S(x_1-x_2)\gamma^5\big]\; .\label{eq:chi1}
\end{eqnarray}
If we now substitute for the quark propagator the expression $S_{\rm full}$, we obtain the formula for the amplitude $\chi_P$ in all instanton sectors in question: $0,\pm1$ (recall that we now have to do with an object billinear in quark fields).
We now pass to the CM and relative coordinates: $x_1=X+x/2$ and $x_2=X-x/2$, and introduce a  wave function by separating the uniform motion of the bound particle as a whole, from the internal motion: $\Phi_P(x)=e^{iPX}\chi(x_1,x_2)$. Finally we obtain:
\begin{eqnarray}
\Phi_P(x)=&&\!\!\!\!\!\!-2\sqrt{\pi}S(x)\gamma^5\sin(Px/2)\nonumber\\*
&&\!\!\!\!\!\!+\frac{\mu}{2\sqrt{\pi}}e^{\gamma_E}e^{ig^2\beta(x)}e^{-i\theta\gamma^5}\gamma^5\cos(Px/2)\; .\label{eq:phip}
\end{eqnarray}
This amplitude wave function entirely agrees with that found in~(\ref{eq:phi1}) from the factorizable part of the bound state residue.

\section{Summary}
\label{sec:sum}
In conclusion we would like to recapitulate our main results. We have found analytically the polar contribution to the $q\overline{q}$ Green's function. In the topologically trivial case, the residue turns out to be factorizable, which is commonly taken for granted. Out of this we obtained the formula for the Bethe-Salpeter wave function. The situation turned out to be different, if one considers one or two instanton background. We have obtained expression for the residue in this case and found, that the BS function has been modified by the $\theta$-vacuum structure. The full polar term may now be written as a sum of several factorizable terms. One sees that, apart from the formation of the bound state (Schwinger boson), there appears also another way for a pair $q\overline{q}$, not bearing meson quantum numbers, to travel over the space. They may jump from one instanton to the other. This phenomenon is possible in there are at least two instantons.

Moreover if one takes into account also these components, that are usually neglected for $P^2\rightarrow \mu^2$ as being small in comparison with that revealing a pole, one finds another type of singularity --- a branch point connected with the appearance of logarithmic and dilogarithmic functions. They arise because of long-range correlations in the condensed vacuum.

All these instanton contributions vanish, if we consider the $S$-matrix elements, i.e. only transition amplitudes for the `physical' asymptotic states. These, in the Schwinger Model, are Fock states of one or more bosons. The four-point (quark) function is then replaced with the current-current function, for which it is shown, that simple pole structure as well as factorization of the residue reappear. The analytical structure of the Green's function may then be more intricate than that of the $S$-matrix, especially for topologically nontrivial theory.

\section*{Acknowledgments}
I would like to thank to Professors J\'ozef Namys{\l}owski and Krzysztof Meissner for elucidating discussions.

\end{document}